# A Coherent Distributed Grid Service for Assimilation and Unification of Heterogeneous Data Source

Tanvir Ahmed[*]    Mohammad Saiedur Rahaman[**]    Mohammad Saidur Rahman[Ψ]    Manzur H. Khan[¥]

*Abstract* — Grid services are heavily used for handling large distributed computations. They are also very useful to handle heavy data intensive applications where data are distributed in different sites. Most of the data grid services used in such situations are meant for homogeneous data source. In case of Heterogeneous data sources, most of the grid services that are available are designed such a way that they must be identical in schema definition for their smooth operation. But there can be situations where the grid site databases are heterogeneous and their schema definition is different from the central schema definition. In this paper we propose a light weight coherent grid service for heterogeneous data sources that is very easily install. It can map and convert the central SQL schema into that of the grid members and send queries to get according results from heterogeneous data sources.

## 1. Introduction

Grid computing is the arrangement of computer resources from various administrative domains dedicated to a common task, usually to a technical, scientific or business problem that requires a great number of computers processing cycles [13]. Grid computing uses middleware to synchronize disparate IT resources across a network that allows them to function as a virtual whole. The goal of a computing grid system is like that of the electrical grid which is to provide users with access to the resources they need, when they need them [12].

---
[*] Master of Science Student, Department of Computer Science
American International University Bangladesh (AIUB) Kemal Ataturk Avenue, Dhaka, Bangladesh, Email: - tanvir@aiub.edu

[**] Master of Science Student, Department of Computer Science
American International University Bangladesh (AIUB) Kemal Ataturk Avenue, Dhaka, Bangladesh, Email:- saied@aiub.edu

[Ψ] Master of Science Student, Department of Computer Science
American International University Bangladesh (AIUB) Kemal Ataturk Avenue, Dhaka, Bangladesh, Email:- saidur@aiub.edu

[¥] Assistant Professor, Department of Computer Science
American International University Bangladesh (AIUB) Kemal Ataturk Avenue, Dhaka, Bangladesh, Email:- manzur@aiub.edu

One of the key strategies of grid computing is using software to divide and assign pieces of a program in the midst of several computers, sometimes up to several thousands. Grid computing is distributed and considered as large-scale cluster computing, as well as a special form of network-distributed parallel processing [11].

Grid computing mechanism works by distributing computational resources. But it maintains a central controller of the process. A central server acts as a central controller or team leader. It also performs traffic monitoring tasks. This controlling cluster server divides a task into subtasks, and then assigns the work to computers with surplus processing power on the grid. Additionally it also monitors the processing and the subtask routines. In case a subtask routine fails the controller restarts or reassigns it. When all the subtasks are completed, it is the responsibility of the controlling cluster server to aggregate the results and move towards the next task until the entire job is completed.

In a global grid, it is certain that machines are placed on many different networking areas and on the Web. Because of their processing in so many different purposes, network latency can be a problem. Again, before any type of processing can occur, the available resources must be identified and located accordingly. Access to them has to be negotiated, and the used hardware and software must be configured to effectively use the resources, which often are many smaller computers [10].

A layer of middleware is used by the Grids to communicate with and manipulate heterogeneous data sets and hardware. In some fields like astronomy, hardware cannot reasonably be moved and is prohibitively costly to replicate on other sites [12].

Grid services are heavily used in database fields. Various types of techniques are available to gather data from distributed database. Most of them follow the Open grid service architecture (OGSA) [9]

Nowadays grid services are extensively used for data collection. They are also appropriate for high data intensive applications. For vast amount of data processing, a grid service can be the best choice. But usually most of them are designed for a specific purpose or specific type of application. For example, they are highly used for bioinformatics data collection and processing, specific type of data mining, various market analysis, distributed database management etc. In general they follow common schema and homogeneous data sources. But it is obvious that there will be situations where grid services will be required to work on heterogeneous database management systems. In such cases commonly available grid services will not work as they require common schema definition. On the other hand if the members of the grid have different schema definition, then they have to create their own web service to distribute data. It certainly would create complexity for the members inside a grid and reduce interoperability immensely. Thus the objective of grid service would be at stake.

In section 2 of this paper, an overview of the current trend of the technology is presented. Our proposal of a grid service for assimilation and unification of heterogeneous data Source eliminating the existing problems is explained in section 3. We have implemented our idea to test that it really meets the need of assimilating and unifying heterogeneous data sources with simulated data. This is illustrated in section 4. Section 5 explains the benefits of our model and gives the idea for the future advancement. Finally the paper concludes in section 6.

## 2. Background Study & Related Works

Various techniques of grid services are available to collect data from distributed databases. Most of the techniques follow the Open grid service architecture (OGSA DAI) [2] [3] [5] [8] [9]. The Open Grid Services Architecture (OGSA) illustrates the architecture for a service-oriented grid computing environment developed within the Global Grid Forum (GGF). OGSA is based on several other Web service technologies, especially SOAP and WSDL, but it targets to be largely agnostic in relation to the transport-level handling of data [9]. To access and control data sources and sinks, The Open Grid Services Architecture – Data Access and Integration (OGSA-DAI) project is developing an efficient Grid-enabled middleware implementation of services and interfaces [8]. For the stage of the project just completed, these data sources and sinks were restricted to be relational and XML database management systems (DBMS). But the main purpose for which the framework has been designed is to allow other data sources such as file systems to be accessed through the same interfaces.

Paper [7] examines a way to integrate grid environment for warehousing distributed heterogeneous relational data where the data are physically located at many sites. They used the existing grid middleware concept to create a data warehouse where data are collected from heterogeneous DBMS situated at different sites. For data collection they used web services at different sites which feature an easy access for web applications.

In case of bioinformatics research grid service is widely used for data collection from different locations [4] [6]. In [4] a grid service is developed to collect bioinformatics information from different data sources. They used web services and used RMI to collect data from different bioinformatics database.

In [1] they put forward a Gird-Based Integration Model (GBIM) for uniformly accessing heterogeneous database systems. They used web services and OGSA to integrate heterogeneous data sources. Moreover they also used a special type of query to request data from different sources.

Some of the above systems use similar data sources and others use heterogeneous data sources. But those using heterogeneous data sources are designed for specicific applications. Additionally they use webs services at different sites which need extra development effort and make installation much complex. Moreover these existing systems use different application dependent commands to gather data and to ensure heterogenity. This is the motivation of our research to develop an application independent grid service using the existing SQL to ensure assimilation and unification of heterogeneous data.

## 3. Proposed Model

The proposed system consists of two main units as described below:

### 3.1. Centre Unit

Centre unit is the coordination part of this system. Actually the need of grid service is realized and implement by the centre. The grid service is mainly for centre's requirement. For example, University Grant Commission (UGC) wants to create portal where the information of all the universities will be available. So UGC can provide this grid service to all the universities and tell them to install it. Here the UGC will be the centre and all the universities will be the grid members. Any user will be able to request data from the UGC portal but the main service will be provided by the different universities. The UGC portal will distribute the query to all the university and after getting the results from different universities the UGC portal will display them. So here the UGC is the grid centre. The components of a Centre are given below:

*3.1.1. Virtual Database*

The virtual database is nothing but an XML file. This XML file contains the centre's database schema that is, the table names and column names which will be used in the centre's SQL. The centre actually does not contain any database. To write SQL, centre follows the schema definition from this XML file. Centre assumes that it has a database with the schemas that are available in this XML. This XML also contain the centre's network address, which is the combination of IP address and port address of the centre register server. The following centre.xml shows a sample virtual database. Details of different tags and attributes are given below:

- Table tag: Contains the tables information
  - Name (Attribute) : Contains the table name
  - Description (Attribute) : Contains the description of the table
  - Column (Child tag): Contains information about column of the table
    - Name (Attribute): It contains the name of the column.
    - Description (Attribute): Contains the description about the column
    - Data type (Attribute): Contains the data type of the column

*Centre.xml*

```xml
<?xml version="1.0" encoding="UTF-8" ?>
<database>
  <table name="student" description="Student Information">
    <column name="studentid" description="Stuent IDentifier" datatype="string" />
    <column name="studentname" description="Student's Name" datatype="string" />
    <column name="departmentid" description="Department IDentifier" datatype="string" />
    <column name="sex" description="Sex" datatype="string" />
    <column name="CGPA" description="CGPA" datatype="float" />
  </table>
  <table name="Department" description="Department Information">
    <column name="departmentiden" description="Department IDentifier" datatype="string" />
    <column name="departmentName" description="Department Name" datatype="string" />
    <column name="TotalCredit" description="NumberOfCredit" datatype="float" />
  </table>
</database>
```

### 3.1.2. Centre Register Service

Centre register service works as a register server. It always listens to a port of the centre machine. It reads and stores new member's information in an XML file. The name of the xml is "Gridlist.xml". It contains three data of the grid member.
- Member Name
- Member Network Address
- Member Port Number

The following xml shows an example of storing grid members list.

```xml
<?xml version="1.0" standalone="yes" ?>
<GridList>
  <Grid gridOrganization="temp" GridNetworkAddress="127.0.0.1" Port="8888" />
  <Grid gridOrganization="test1" GridNetworkAddress="172.16.43.13" Port="8111" />
  <Grid gridOrganization="test2" GridNetworkAddress="172.16.43.13" Port="8222" />
  <Grid gridOrganization="test3" GridNetworkAddress="172.16.43.13" Port="8333" />
</GridList>
```

When the centre register server starts it reads the GridList.xml file and displays the existing grid members list. After displaying the list it starts listening to the port for new member registration. If it gets a data in the port then it checks whether this member is already available or not. If the member already exists, then it updates the information of the member in the GridList.xml file. But if the member is new then it adds this member in the grid list and adds this information in the GridList.xml file. After adding or updating the member it again display the member list and start listening to the port. This process is continued until the grid register service is stopped. Figure 1 shows the overview of the steps of grid register service.

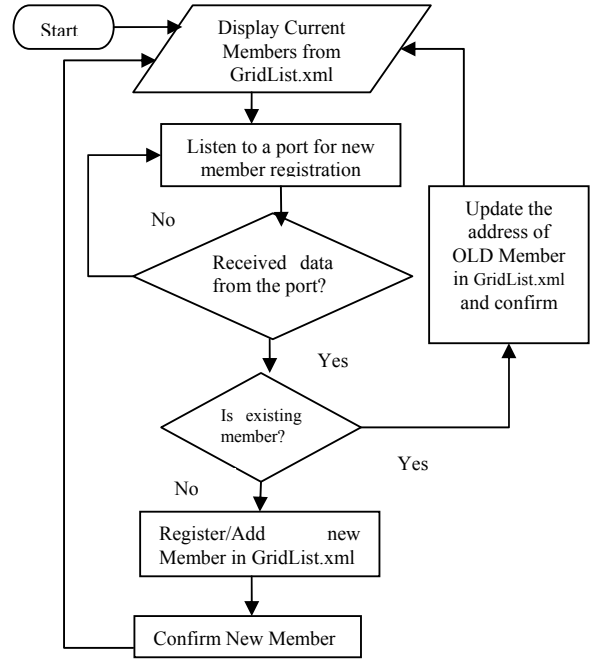

**Figure 1: Overview of the steps of Grid Register service**

### 3.1.3. Grid consumer application / Portal

This component is the consumer of the grid service. All the grid services that have been designed are to facilitate this application. An application or portal can be used to consume the services of the grid members. This application can take a raw SQL SELECT statement and can distribute this SQL to the grid members to execute it. This should be done in multi threads so that it can be executed in parallel. For high level user proper user interface can be provided to consume the services of the grid members. For example, in a portal of UGC a user can search for information about 'CSE' department. Then the SQL starts working in the background of the portal. This SQL is distributed among the grid members. After getting the results from the grid members it synchronizes the results and displays it to the user. This is then displayed to the user in formatted manner. The user will think that he/she is getting data from the UGC portal or application.

## 3.2. Grid member unit

This unit actually performs the execution of the query. Grid members register themselves with the centre. Grid services are installed in these member's machines. A Grid Service negotiates with the centre for providing services. Details of its important components are described in the following sections.

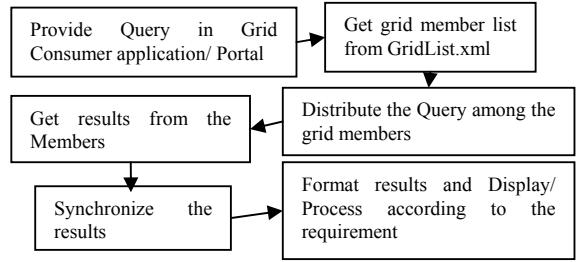

**Figure 2: Overview of the steps of Grid consumer application/Portal**

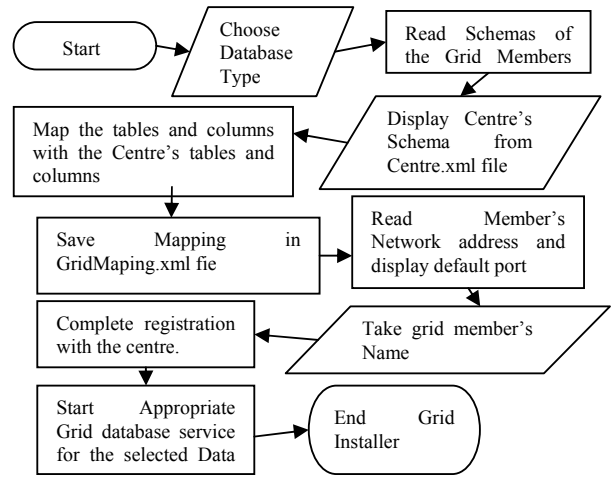

**Figure 3: Overview of Grid service installer tasks**

### 3.2.1. Grid Service Installer

This tool is used to install the grid service in the grid member's machine. This is a very user friendly installer and most of the tasks are automated. The centre's schema definition file which is called 'centre.xml' is provided with the installer. The grid members may have different types of DBMS. So when the installer is started the person who is installing has to choose the DBMS where the data are stored. After choosing the appropriate DBMS the installer will help the person to choose the database, tables and columns. This will also display the centre's tables and columns information with proper description. This information is taken from the centre.xml file. The users have to map its tables and columns with the centre's tables and column. After completing the mapping an XML file will be created called 'GridMapping.xml' where the mapped information will be saved. After that the installer will display a registration page. In this page the grid member/ organization name, network address and port number have to be provided. The IP address of the current computer is automatically read by the installer. It also shows a default port address. But the person who is installing the grid has freedom to change it. After registering a confirmation will be provided to the person and then the grid service for that database will be automatically started by the installer. Figure 3 displays the overall steps of the grid service installer.

### 3.2.2. Database grid service

This is the actual service that acts as a grid service in this system. This service runs in each grid member's machine. This data grid service is multi thread program. That means it can receive multiple request simultaneously. It does the following tasks:

- Takes SQL from the centre
- Maps the SQL with the grid member's Schemas
- Regenerates new SQL from old SQL
- Sends the new SQL to the grid member's DBMS
- Takes the result from the grid member's DBMS
- Sends the result to the Centre.

This service always listens to the port that was given at the time of installation of the grid service. This port number is available in the GridMapping.xml file. When it gets any SQL through the port it receives it and completes the mapping operation. The mapping information is available in the GridMapping.xml file. After mapping, it generates a new SQL for the grid member that can be understood by the gird member's DBMS. The new SQL is then sent to the member's DBMS. The DBMS then execute the SQL. The service takes the result from the DBMS and sends the result to the centre.

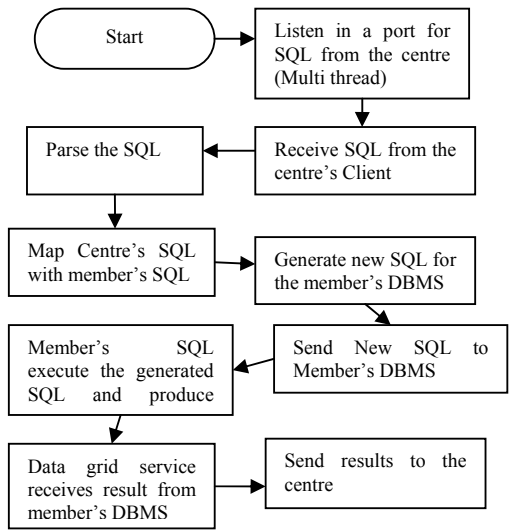

**Figure 4: Overview of data grid service's tasks for each member's grid service**

### 3.2.3. Query parser and mapping module

Mainly this module falls inside the data grid service part. In this portion the received SQL SELECT statement is parsed and the tables and columns of each clause are identified. In an SQL SELECT statement various types of clause are available like Select lists, Joining, group by, where, functions, order by and there may be sub queries. All these clauses may contain different column names and table names. In this portion all of these columns and table names are identified. After identification the table names and column names, it is mapped and replaced by the grid member's column names. After replacing each tables and columns of each clause, it produces a new SQL that is appropriate for the member's schemas and DBMS.

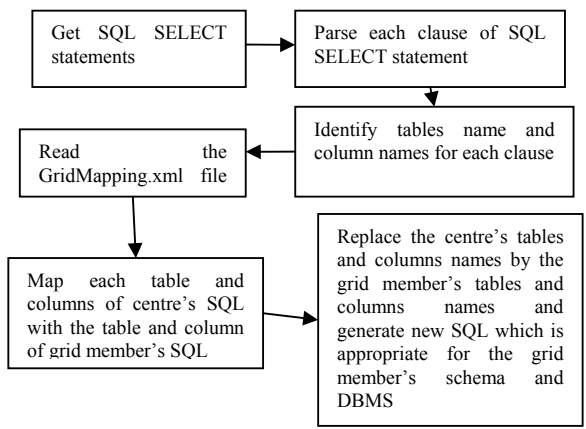

**Figure 5: Overview of Query Parser and Query mapping module**

This component take help from the GridMapping.xml file for mapping the centre's query with the grid member's query. A sample GridMapping.xml is given below.

```xml
<?xml version="1.0" standalone="yes" ?>
- <Mapping>
    <ConnectionString value="Data Source=tanvir;Persist Security
      Info=True;User ID=scott;password=tiger;Unicode=True" />
  - <Table CenterTable="student" GridTable="STUD">
      <Column CenterColumn="studentid" GridColumn="STUID" />
      <Column CenterColumn="studentname" GridColumn="STUNM" />
      <Column CenterColumn="departmentid" GridColumn="DEPTNUM" />
      <Column CenterColumn="sex" GridColumn="SEX" />
      <Column CenterColumn="CGPA" GridColumn="CGPA" />
    </Table>
  - <Table CenterTable="Department" GridTable="DEP">
      <Column CenterColumn="departmentiden" GridColumn="DEPNUM" />
      <Column CenterColumn="departmentName" GridColumn="DEPNAME" />
      <Column CenterColumn="TotalCredit" GridColumn="TOTALCREDIT" />
    </Table>
  - <PortAddress>
      <Port>2222</Port>
    </PortAddress>
  </Mapping>
```

The description of each tag and attributes are given below.

- Connection String (tag): Contains the database connection information, type of database, authentication information, database location etc.
- Value (attribute): Contains the connection string.
- Table (tag): Contains the tables information
- Centre table (attribute) : Contains the name of the centre table
- Grid table (attribute): Contains the name of the current grid member table that mapped with the above centre table.
  - Column (tag) : Contains the column information of the above table
  - Centre Column (attribute): Contains the name of the above centre's table column.
  - Grid Column (attribute): Contains the name of the current grid member's column of the above table that mapped with the above column of the centre.

### 3.2.4. Grid Member's DBMS

The main target of this grid service is data collection. Grid members have different DBMS installed. It can be Oracle, MS SQL Server, MS Access, My SQL, DB2, Post GRE etc. Each grid member's database is independent of each other. Their database schemas are also different. Regardless of DBMS and Schema, this system can execute the centre's given query. The DBMS does the actual query execution after receiving the new generated query from the mapping module.

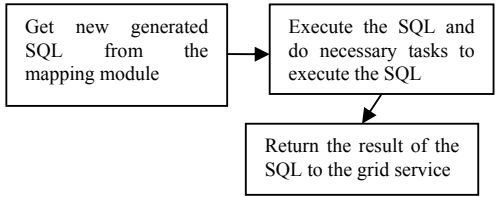

**Figure 6: overview of tasks performed by the grid member's DBMS**

### 3.2.5. Heterogeneity Design

To provide heterogeneity we used the xml file GridMapping.xml. While installing the grid service the person provided the data sources and schemas information. The connection string is generated dynamically by the installer at that moment. This connection string is saved in the Gridmapping.xml file. Because the GridMapping.xml file also saves the mapping information the system becomes schema independent. We have created different grid services for different DBMS especially for some popular database management systems like MS SQL Server, Oracle, My SQL, MS Access. In addition we have also created grid service for ODBC data source. As a result if a database management system does not fall in the above lists then the data source can be added in the ODBC data source. As a result this will work for any database management system which supports SQL.

The proposed services in this paper work in such a way that it takes the database connection information from the GridMapping.xml file along with the schema definition and as a result this system works for any type of database.

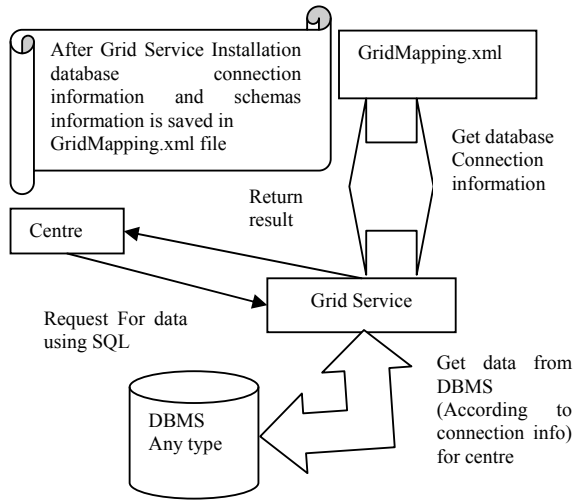

**Figure 7: Heterogeneity design**

### 3.3. Overall Conceptual Design

The centre has its own schema definition of the database without having ay actual database. The schema definition is stored in a simple xml file. In the grid consumer software if a user writes any SQL query for the centre schema then this query will be transferred to all of the grid members. Interesting thing is that the grid member's databases are heterogeneous and their schema definition is different from the centre's schema definition. So the grid service will map and convert the centre's SQL into the grid members SQL and send the query to the DBMS to get the result. Each grid members will do the same operation and will send the result of the query to the centre. The centre will take the results and present the results to the user who requested the query. The user even doesn't know that this query is processed in different sites and the result is coming from different sites. Here the main processing is done by the grid members.

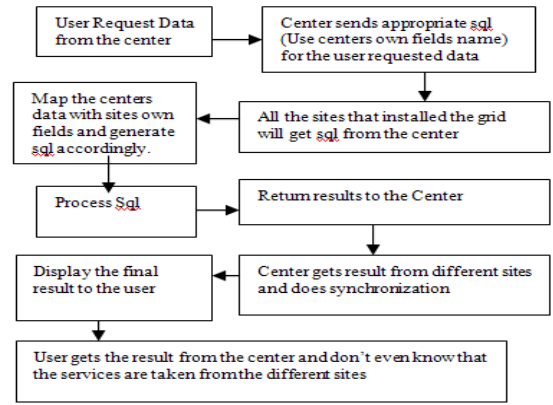

**Figure 8: Overall Conceptual task of the system**

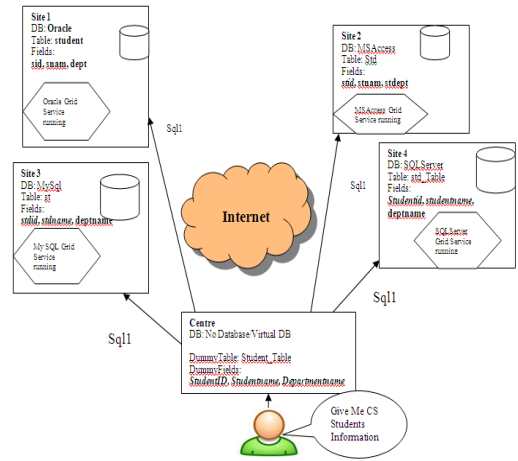

**Figure 9: Example scenario of the system**

## 4. Simulated Output of Proposed Scheme

We have generated some simulation results by our developed application based on our proposed scheme. The following is an example that depicts a clear idea of the words presented here.

Figure 10 shows an example of gird register service. Here we can see that the centre register service is running and displaying the current grid members from the GridList.xml. It also listening to a port and waiting for new member registration.

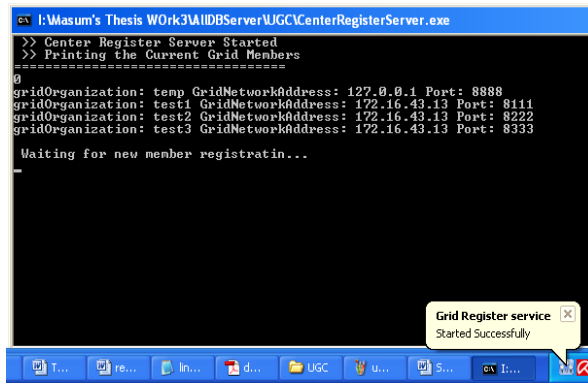

**Figure 10: Grid register service running.**

Figure 11 shows an example of grid installer. After choosing MY SQL from the database list, it displayed this form to perform the mapping and store mapping information.

Figure 12 shows an example of grid consumer software and the grid service that is running in different servers. In the SQL field the user wrote a complex SQL and the result field displays the results that were received from the members of the grid. It also shows that after receiving the SQL, the grid service regenerated new SQL depending on its own schema.

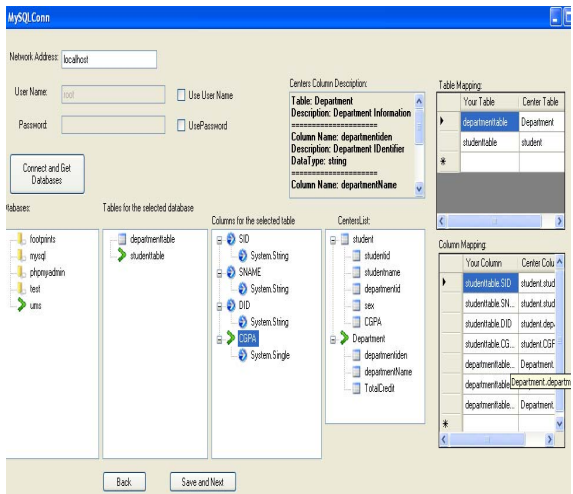

**Figure 11: An example of grid installer**

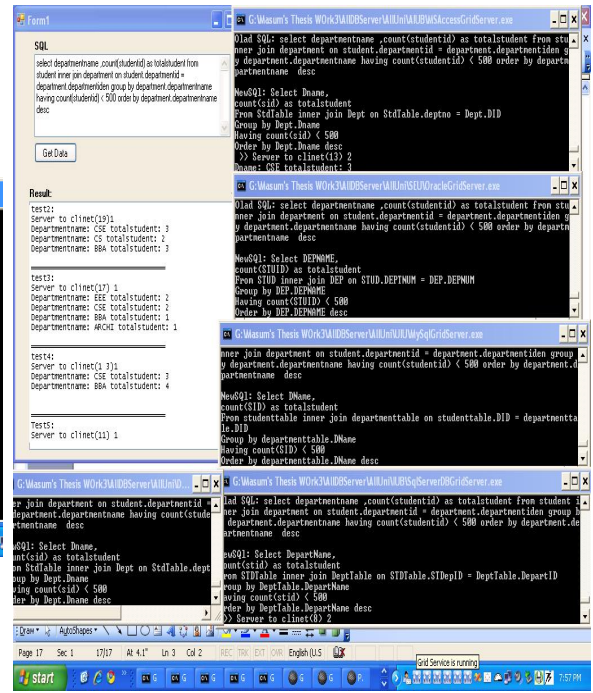

**Figure12: Example grid service consumer software**

## 5. Benefits & future works

There are many centre based application that required huge amount of data processing. Sometimes installation of such type of system is difficult due to heterogeneity of database management system in different sites. Another additional difficulty is if each site has a different schema, tables and columns. In that case each site has to develop its own web services by its own. In this paper a coherent distributed grid service is designed which is able to assimilate and unify the heterogeneous data sources. As a result any centre based application can use this system. This system works in a situation where the database management systems and schemas are also heterogeneous. Here all the sites have their own individual systems and databases. They just have to install the grid service to their system. The installation process is very simple and user friendly. After installing the grid service the centre can consume the service using a portal or application to get data from all the member's sites. Some features of our proposed service are discussed below.

- Heterogeneous DBMS support: This grid service can run on any type of database management system that supports SQL. This system highly supports the following popular DBMS: Oracle, My SQL, MS Access, MS SQL Server, etc. For other type of DBMS it uses ODBC data source. If a data source is added in the ODBC data source then our system can access the data of that data source.

- Heterogeneous Schema support: This system can operate on heterogeneous schema. That means if the grid members have different table name and column name the system will work in such situation. Actually it does a mapping of the centre SQL to the Grid member's SQL.

- Distributed Query Processing: As this is a grid service, the tasks of the centre is distributed among the grid members. As a result parallel processing is performed in different machine. Here the SQL of the centre is distributed among the grid members. The grid members do the SQL mapping and execution and then return the result to the centre. The centre collects all the results from the grid members and does further processing or displaying.

- Transparent to the user: This system is completely transparent to the users. The users request queries to the centre and the centre perform the operation by the grid members. Finally the results are displayed by the centre. But the users do not know that the operations are performed by the grid members. They will think that the results are coming from the centre.

- Low resource requirement: Generally grid services are very resource hungry. But this system is very light weight and requires very low resources to operate. Especially the centre computer can be a very simple computer or device and it does not require any database.

- Easy To Install: This grid service is very easy to install. Generally installation of grid services is very complex and requires computer or database administrator. But this installer is very user friendly and most of the tasks are performed automatically.

- Implementation in any type of organization: This grid service can be implemented in any type of centre based organization. For example the stock exchange can use this grid service to collect information from different companies or the public library can use this system to search books in different libraries etc.

More research can be done for providing extra intelligence so that it can detect similar type of columns and tables for mapping automatically. Another issue can be the SQL conversion for different database management systems. Extra security can be imposed so that unauthorized data can't be accessed.

## 6. Conclusion

In this paper we presented a coherent distributed grid service which is designed for assimilation and unification of heterogeneous data sources. This is an excellent idea for a centre based system to collect data from independent sites those have different schema and heterogeneous database management system. As the processing is distributed among sites, the centre can be a very simple low processing computer. Most of the information is stored in XML. As a result this is a light weight service. Another important feature is that the grid service installation is very simple. Here the grid member doesn't have to create their own web services. They just have to install this grid service to provide service to the centre. The general public users will be benefited from this system if the centre of different organizations uses this system to develop their portal. The example of the centre can be UGC, Public Library, Stock exchange and different government agencies that coordinate with different types of organization or matters. In future we aim to convert SQL for different database management systems and impose extra security to restrict unwanted access to data. Currently we are also trying add intelligence to detect similar types of columns and tables on different sources for mapping.

## Biographies

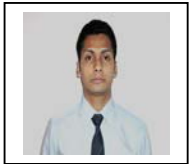

Tanvir Ahmed earned his B.Sc. in Computer Science & Engineering in 2008. Now he is pursuing his M.Sc. in Computer Science in the American International University – Bangladesh (AIUB). Currently he is working as a Lecturer in Department of Computer Science in the same university. His research interests include grid computing, and data mining.

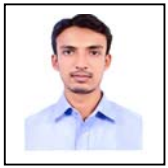

Mohammad Saiedur Rahaman earned his B.Sc. in Computer Science & Engineering in 2007. Now he is pursuing his M.Sc. in Computer Science in the American International University – Bangladesh (AIUB). Currently he is working as a Lecturer in Department of Computer Science in the same university. His research interests include wireless mesh networks, wireless sensor networks, grid computing, mobile & multimedia networks and data mining.

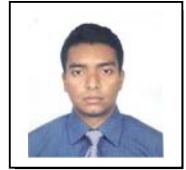

Mohammad Saidur Rahman earned his B.Sc. in Computer Engineering in 2006. Now he is pursuing his M.Sc. in Computer Science in the American International University – Bangladesh (AIUB). Currently he is working as a Lecturer in Department of Computer Science in the same university. His research interests include grid computing, wireless mesh network, video on demand.

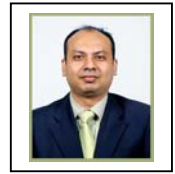

At present, Manzur H Khan is an Assistant Professor of the Department of Computer Science of the American International University-Bangladesh (AIUB). Initially, he completed his Bachelor and Master of Science Degree in Physics from Dhaka University in 1999. After that he completed his Master of Science in Business Information Technology from the University of Westminster, UK in 2001. Manzur joined AIUB as a Faculty of the Department of Computer Science in 2002 and has been teaching numerous undergraduate and postgraduate courses since then. He was also appointed as the Coordinator of the Office of Student Affairs as an additional administrative position in 2005. Recently, he has been promoted as the Director of Office of Student Affairs of AIUB. Manzur's fields of interest in Computer Science and Information technology include Software Engineering, Object Oriented System Analysis and Design, System Analysis and Design, Database Management Systems, and MIS and Decision Making Technique.